\documentclass[a4paper,11pt]{article}
\pdfoutput=1 

\usepackage{jheppub} 

\usepackage[T1]{fontenc} 
\usepackage{enumitem}

\usepackage{tikz}
\usetikzlibrary{decorations.pathmorphing}
\usepackage{subcaption} 
\usepackage{caption}
\usepackage{tkz-tab}
\usepackage{amsmath}
\usepackage{amssymb}
\usepackage{amsfonts}
\usepackage{amsthm}
\usepackage{mathrsfs}
\usepackage{comment}
\usepackage{graphicx}
\usepackage{epstopdf}
\usepackage{empheq}
\usepackage{hyperref}
\usepackage{multirow}

\def\be {\begin{equation}}
\def\ee {\end{equation}}
\def\bea {\begin{eqnarray}}
\def\eea {\end{eqnarray}}
\def\nn {\nonumber}
\def\cL{{\cal L}}

\def\gsim{\, \rlap{$>$}{\lower 1.1ex\hbox{$\sim$}}\,}
\def\lsim{\, \rlap{$<$}{\lower 1.1ex\hbox{$\sim$}}\,}

\definecolor{purple}{rgb}{0.7,0,1}

\title{\boldmath Self-similar solutions and critical behavior in Einstein-Maxwell-dilaton theory sourced by charged null fluids}

\author[a]{Pedro Aniceto}
\author[b,1]{Jorge V. Rocha,\note{Corresponding author.}}

\affiliation[a]{Center for Mathematical Analysis, Geometry	and Dynamical Systems,\\
 Instituto Superior T\'{e}cnico, Universidade de Lisboa,\\
  Av. Rovisco Pais 1, 1049-001, Lisboa, Portugal}
\affiliation[b]{Departament de F\'isica Qu\`antica i Astrof\'isica, Institut de Ci\`encies del Cosmos (ICCUB),\\
Universitat de Barcelona, Mart\'i i Franqu\`es 1, E-08028 Barcelona, Spain}

\emailAdd{pedro.aniceto@tecnico.ulisboa.pt}
\emailAdd{jvrocha@icc.ub.edu}

\abstract{
We investigate continuously self-similar solutions of four-dimensional Einstein-Maxwell-dilaton theory supported by charged null fluids. We work under the assumption of spherical symmetry and the dilaton coupling parameter $a$ is allowed to be arbitrary.
First, it is proved that the only such vacuum solutions with a time-independent asymptotic value of the dilaton necessarily have vanishing electric field, and thus reduce to Roberts' solution of the Einstein-dilaton system.
Allowing for additional sources, we then obtain Vaidya-like families of self-similar solutions supported by charged null fluids. By continuously matching these solutions to flat spacetime along a null hypersurface one can study gravitational collapse analytically. Capitalizing on this idea, we compute the critical exponent defining the power-law behavior of the mass contained within the apparent horizon near the threshold of black hole formation. For the heterotic dilaton coupling $a=1$ the critical exponent takes the value $1/2$ typically observed in similar analytic studies, but more generally it is given by $\gamma=a^2(1+a^2)^{-1}$. The analysis is complemented by an assessment of the classical energy conditions.
Finally, and on a different note, we report on a novel dyonic black hole spacetime, which is a time-dependent vacuum solution of this theory. In this case, the presence of constant electric and magnetic charges naturally breaks self-similarity.
}

\begin{document} 
\maketitle
\flushbottom


\section{Introduction}
\label{sec:Intro}

Gravitational collapse is the classical mechanism responsible for the formation of black holes in the universe.
It occurs when the matter density is so high that gravitational attraction, the weakest of all known forces, dominates over all other forces of nature.
It has been understood long ago that stellar-mass black holes can be formed when sufficiently massive stars exhaust their nuclear fuel~\cite{Oppenheimer:1939ue}.
Alternatively, primordial black holes may have formed in the early universe as the result of the gravitational collapse of cosmological density fluctuations~\cite{Hawking:1971ei}, although there is currently no observational evidence for their existence.

Once gravitational collapse sets in, typically it is not guaranteed that a black hole will form.
This only occurs if the initial conditions are sufficiently `strong' in some sense.
Otherwise, the matter contracts without ever generating sufficiently high densities to produce an event horizon and then fully disperses, leaving behind flat space.
This problem was first investigated by Christodoulou in a series of papers~\cite{Christodoulou:1986zr,Christodoulou:1987vv}, taking a massless scalar field as the matter model. 
At the threshold between forming or not a black hole during gravitational collapse, critical phenomena arises.
This was revealed by Choptuik in a seminal paper~\cite{Choptuik:1992jv}, which studied numerically the same problem previously considered by Christodoulou and which set much of the groundwork for the field that became known as {\em critical collapse} (see~\cite{Gundlach:2007gc} for a review).

Near the threshold of black hole formation Choptuik found universal behavior, emergent self-similarity, and the famous power-law scaling for the mass of the black hole formed,
\be
M_{BH} \propto (p-p_*)^\gamma\,.
\label{eq:Choptuik}
\ee
The so-called critical exponent $\gamma$ is universal in the sense that it is independent of the details of the initial conditions leading to black hole formation. The quantity $p$ parametrizes the initial condition and $p_*$ is its critical value, which is not universal.
Besides being a problem of interest in its own right, critical collapse is relevant for studies of primordial black holes~\cite{Niemeyer:1999ak}, and recently has been considered also from a holographic perspective~\cite{Chesler:2019ozd}.

The main goal of this paper is to investigate critical collapse in the context of Einstein-Maxwell-dilaton (EMD) theory by employing continuous self-similar (CSS) solutions. This theory can be thought of as comprising a continuum family of models, each one being determined by a coupling constant ---the dilaton coupling $a$ that controls the strength of the coupling between the gauge field and the scalar field. The theory can be regarded simply as a generalization of the Einstein-massless scalar model, but for discrete values of the dilaton coupling it naturally arises from four-dimensional low-energy effective string theories. Critical collapse has been intensively studied for a multitude of matter models~\cite{Gundlach:2007gc}, but it was only recently that it was addressed in the context of the EMD system~\cite{Rocha:2018lmv}.

The techniques we adopt are entirely analytic and are based on an early approach developed by Brady~\cite{Brady:1994xfa} for the Einstein-massless scalar model. The basic idea is to construct a global spacetime that describes gravitational collapse by gluing CSS solutions (with a non-trivial scalar field) onto flat spacetime. The case of continuous self-similarity ---as opposed to discrete self-similarity--- is the most favorable for deriving such solutions analytically. For the Einstein-scalar system, source-free CSS solutions were obtained long ago by Roberts~\cite{Roberts:1989sk}, and Brady employed this 1-parameter family of solutions to investigate critical collapse analytically.

In a previous paper~\cite{Rocha:2018lmv} continuously self-similar collapses in EMD theory were studied, but it focused exclusively on source-free solutions. Moreover, it was found that regularity at the origin implies the vanishing of the Maxwell field over the entire space. In fact, we will prove that the only such {\em vacuum} solutions of the EMD model with a time-independent asymptotic value of the dilaton necessarily have vanishing electromagnetic field, under the assumptions of spherical symmetry and continuous self-similarity. Therefore, the natural setting to consider is the introduction of sources in the field equations, without abandoning the convenient assumption of CSS.

We should note that for the Einstein-Maxwell-dilaton system some spherically symmetric, non-static, source-free solutions are explicitly known (for dilaton coupling $a=1$ only)~\cite{Gueven:1996zm, Aniceto:2017gtx}, but they are not self-similar. In fact, they could not possibly be self-similar: any constant non-vanishing (electromagnetic) charge ---as is the case--- will introduce a dimensionful quantity that precludes any scale invariance of the system.

However, if we allow for sources, one can indeed find CSS solutions in EMD theory. These are relatively simple to obtain for the case $a=1$ by taking the solutions of ~\cite{Gueven:1996zm, Aniceto:2017gtx} and relaxing the constraints that produced vacuum solutions. In particular, allowing the mass parameter $M$, the dilaton charge $D$ and the electric charge $Q$ to all be linear in advanced/retarded (null) time yields a 2-parameter family of self-similar spacetimes supported by charged null dust. Alternatively, one can directly take the well-known GMGHS static solutions~\cite{Gibbons:1987ps,Garfinkle:1990qj} and promote the charges to be functions of the advanced/retarded time coordinate, as first done by Vaidya for pure Einstein theory~\cite{Vaidya:1951zz}. This approach within EMD  theory was used in Ref.~\cite{Aniceto:2015klq} without the assumption of self-similarity.

Given that spherically symmetric static solutions of EMD theory are actually known for generic values of the dilaton coupling~\cite{Gibbons:1987ps,Garfinkle:1990qj}, this procedure can be reproduced for $a\neq1$, although it is technically more involved. From the form of the corresponding static solutions it is expected that the metric components will display non-integer exponents, and this should ultimately be reflected in a non-trivial critical exponent, thus becoming an explicit function of the dilaton coupling $a$. This will be confirmed by extending Brady's matching calculation to the case of the EMD system.

Now, for the standard Einstein-scalar system one gets a very simple result for the critical exponent, namely $\gamma=1/2$~\cite{Brady:1994xfa}. This value is also commonly observed for many other matter models where some form of critical exponent can be computed analytically~\cite{Strominger:1993tt,Peleg:1994wx,Zhang:2014dfa,Wang:2019bof}. A notable exception is found in the case of Brans-Dicke theory~\cite{deOliveira:1995cn,Chiba:1996zu,Yazadjiev:2003bp} where the critical exponent actually differs from $1/2$, although it can be straightforwardly obtained from the Einstein-massless scalar value via a conformal transformation.

What is then the critical exponent for the case of Einstein-Maxwell-dilaton theory? We will find that it can span the full range $[0,1)$, and is given specifically by
\be
\gamma=\frac{a^2}{1+a^2}\,.
\label{eq:formulagamma}
\ee
Interestingly, for the special value of the dilaton coupling corresponding to the heterotic string, $a=1$, one recovers the typical value $\gamma=1/2$.

We will be restricting our investigations to the EMD theory with vanishing cosmological constant; otherwise, the introduction of such a fixed scale into the problem would once again yield an obstruction to self-similarity. Nevertheless, it is worth mentioning that the inclusion of a negative cosmological constant, although adding structure to the black hole formation process, leaves unchanged the Choptuik critical exponent~\cite{Bizon:2011gg}. Moreover, asymptotically anti-de Sitter solutions of the EMD system have been extensively considered for holographic applications to condensed matter physics. In particular, that setup allows for static electrically charged black branes with running scalars. The near-horizon behavior of those solutions displays Lifshitz scaling symmetry~\cite{Goldstein:2009cv}, under which time and spatial coordinates scale differently: $t \to \lambda^z t,\, x_i \to \lambda x_i$. The so-called dynamical exponent $z$ characterizing this symmetry also depends non-trivially on the dilaton coupling $a$, and its inverse $1/z$ has a functional form identical to~\eqref{eq:formulagamma}, up to a multiplicative constant. It is plausible that there is a connection between the dynamical exponent $z$ and the critical exponent $\gamma$ we compute, though we do not explore this avenue in the present paper.

Somewhat marginally related to this whole story, we also report on a novel class of time-dependent dyonic vacuum solutions of EMD theory with $a=1$. These spacetimes break continuous self-similarity by introducing constant electric and magnetic charges. For generic values of the charges the spacetime is fully determined by solving a certain second order non-linear ordinary differential equation, but when the two charges are equal in magnitude we are able to write down explicitly the resulting 2-parameter family of solutions.

The outline of the paper is the following.
Section~\ref{sec:EOMs} defines the Einstein-Maxwell-dilaton theory and specifies the corresponding equations of motion, of which we will be interested in obtaining spherically symmetric and CSS solutions.
In section~\ref{sec:SSconditions} we analyze source-free solutions and prove that Roberts' (uncharged) spacetime is in fact the most general solution within this class, under a well-motivated technical assumption about the action of the homothety on the matter fields.
Section~\ref{sec:CSSheterotic} presents a family of CSS solutions supported by null dust when the dilaton coupling is $a=1$, and details the exact computation of the critical exponent.
More generic couplings are then considered in sections~\ref{sec:genericCSS} and~\ref{sec:CSSsourcedSols}, the former laying down the general framework, and the latter considering specifically a 2-parameter class of CSS solutions supported by null fluids.
Section~\ref{sec:Dyonic} derives the new time-dependent dyonic solutions, before taking stock and making some general remarks in section~\ref{sec:Conc}.
Appendix~\ref{App1} contains the analysis of the energy conditions for the solutions obtained in sections~\ref{sec:CSSheterotic} and~\ref{sec:CSSsourcedSols}.

\section{Equations of motion for EMD theory with sources}
\label{sec:EOMs}

The four-dimensional Einstein-Maxwell-dilaton model we consider is governed by the following action (we adopt Planck units throughout, setting Newton's constant and the speed of light to unity, $G=c=1$),
\be
{\cal S} = \frac{1}{16\pi} \int dx^4\, \sqrt{|g|} \left[R-2(\nabla\Phi)^2-e^{-2 a\Phi}F^2 \right]  + \int dx^4\,\cL_{\rm sources}\,.
\label{eq:action}
\ee
Here $g$ represents the determinant of the metric $g_{\mu\nu}$, $A_\mu$ is the Maxwell field whose field strength is $F_{\mu\nu}=\partial_\mu A_\nu-\partial_\nu A_\mu$, and $\Phi$ is the dilaton. The scalar and vector fields couple with a strength controlled by the so-called dilaton coupling constant $a$.

Special values of the dilaton coupling appear naturally in different contexts. The four-dimensional low-energy effective action for heterotic string theory takes the form~\eqref{eq:action} with $a=1$, while $a=\sqrt{3}$ corresponds to Kaluza-Klein reduction of 5D Einstein gravity on the circle. Einstein-Maxwell theory is recovered by choosing $a=0$, and consistently setting the dilaton to zero.

In order to allow the incorporation of additional sources for the dilaton, Maxwell and Einstein equations, we include in the action above linear couplings of the dilaton to a scalar source $\Sigma$ and of the Maxwell field to a current $J^\mu$, as well as an extra matter Lagrangian $\cL^{(\rm m)}$ which accounts for a general fluid component,
\be
\cL_{\rm sources} = \sqrt{|g|}\left[ \Phi\Sigma+ A_\mu J^\mu \right] + \cL^{(\rm m)}\,.
\label{eq:sources}
\ee
The field equations derived from Eqs.~\eqref{eq:action} and~\eqref{eq:sources} thus read 
\begin{subequations}\label{eq:fieldeqs}
\bea
&& \nabla^2 \Phi + \frac{a}{2}e^{-2 a\Phi} F_{\mu \nu} F^{\mu \nu}  = -4\pi \Sigma\,,\label{dilatonEOM}\\
&& \nabla_\mu \left( e^{-2 a \Phi} F^{\mu \nu} \right) = -4\pi J^\nu\,,\label{MaxwellEOM}\\
&& R_{\mu\nu} - \frac{1}{2}R g_{\mu\nu} = 8\pi  T_{\mu\nu} \equiv 8\pi \left(T_{\mu\nu}^{\rm (dil)} + T_{\mu\nu}^{\rm (EM)} + T_{\mu\nu}^{\rm (fluid)}\right)\,.\label{EinsteinEOM}
\eea
\end{subequations}
The full stress-energy tensor $T_{\mu\nu}$ has contributions from the dilaton, the electromagnetic field, and from the (charged) fluid,
\begin{subequations}
\bea
8\pi T_{\mu\nu}^{\rm (dil)} &=& 2\nabla_\mu\Phi\nabla_\nu\Phi - g_{\mu\nu} (\nabla\Phi)^2 \,,\\
8\pi T_{\mu\nu}^{\rm (EM)} &=& e^{-2a\Phi} \left( 2F_{\mu \sigma} {F_\nu}^\sigma -\frac{1}{2}g_{\mu \nu}F^2 \right)\,,\\
T_{\mu\nu}^{\rm (fluid)} &=& T_{\mu\nu}^{(\rm m)} + g_{\mu\nu}\Phi\Sigma + g_{\mu\nu}A_\sigma J^\sigma - 2 A_{(\mu} J_{\nu)}\,,
\label{eq:FluidTmunu}
\eea
\end{subequations}
where $T_{\mu\nu}^{(\rm m)} = -\frac{2}{\sqrt{-g}} \frac{\partial \cL^{(\rm m)}}{\partial g^{\mu\nu}}$, and the parentheses around indices stand for the symmetric part, $A_{(\mu} J_{\nu)} \equiv \frac{1}{2}\left(A_\mu J_\nu+A_\nu J_\mu\right)$.
We loosely adopt the terminology ``fluid'' to describe the component \eqref{eq:FluidTmunu} of the stress-energy tensor because for all cases we consider it will turn out to be of the form of a null fluid, as we discuss in section~\ref{sec:CSSsourcedSols}.

Generically, both the dilaton and Maxwell fields source the Einstein equations.
We will refer to solutions without any additional sources ---i.e., for which $\Sigma$, $J^\mu$ and $T_{\mu\nu}^{(\rm m)}$ all vanish--- as {\em source-free} solutions or simply {\em vacuum} solutions\footnote{To be more precise these should be called dilaton-electrovacuum solutions.}.

\section{Spherically symmetric, source-free CSS solutions}
\label{sec:SSconditions}

In practice, the assumption of continuous self-similarity is embodied in the existence of a homothetic vector field (HVF) $\xi$ such that \cite{Carr:1998at}
\be
\cL_\xi g_{\mu\nu} = 2 g_{\mu\nu}\,.
\label{eq:SScond}
\ee
For spherically symmetric and source-free spacetimes, the above condition implies that the scalar and Maxwell fields transform under the homothety according to~\cite{Rocha:2018lmv}
\bea
\cL_\xi \Phi &=& -\kappa\,, \label{eq:homo_dilaton}\\
\cL_\xi F_{\mu\nu} &=& (1-a\kappa)F_{\mu\nu} \,.
\label{eq:homo_Maxwell}
\eea
The minus sign in front of $\kappa$ is conventional and is commonly adopted in the literature.
If one relaxes the assumption of spherical symmetry, the most general homothetic transformation of the Maxwell field allows for an additional contribution from the Hodge dual $\star F_{\mu\nu}$ (at least as long as the dilaton has a timelike gradient) ,
\be
\cL_\xi F_{\mu\nu} = (1-a\kappa)F_{\mu\nu} + \widetilde{\kappa} \star\! F_{\mu\nu}\,,
\ee
where $\widetilde{\kappa}$ is a scalar quantity obeying certain restrictions (see Ref.~\cite{Rocha:2018lmv}). However, since we will focus exclusively on spherically symmetric spacetimes we will set $\widetilde{\kappa}=0$.

We will adopt (advanced) Eddington-Finkelstein coordinates, which is convenient to make contact with previously obtained time-dependent vacuum solutions. In such coordinates, without loss of generality\footnote{See for example~\cite{Martel:2000rn} for a justification of why the cross term $g_{vr}$ can be taken to be constant.} the line element takes the form
\be
ds^2 = -A(\zeta)dv^2 + 2 dv dr + r^2 B(\zeta) d\Omega^2\,,
\label{eq:ansatz_metric}
\ee
where the self-similar variable $\zeta\equiv v/r$ was defined. The homothetic vector field for such a line element is
\be
\xi = v \frac{\partial}{\partial v} + r \frac{\partial}{\partial r}\,.
\label{eq:HVF}
\ee

Regarding the dilaton and Maxwell fields, the homothetic conditions (\ref{eq:homo_dilaton}--\ref{eq:homo_Maxwell}) imply that these fields take the form
\be
\Phi = \phi(\zeta)-\kappa \log v\,,
\qquad
F= -\frac{Q(v,r)}{r^2} dv\wedge dr = -\frac{\zeta^{1-a\kappa}\overline{Q}(\zeta)}{r^{1+a\kappa}} dv\wedge dr\,.
\label{eq:ansatz_matter}
\ee
This is the most general ansatz for the matter fields consistent with continuous self-similarity and with spherical symmetry.
In writing~\eqref{eq:ansatz_matter} we are taking the Maxwell field to be purely electric, otherwise the field strength could also contain a magnetic piece, $P\sin\theta d\theta\wedge d\varphi$.
Although the homotheticity parameter $\kappa$ can take any real value in principle, we will fix $\kappa=0$ for the rest of the manuscript, so that the large-$r$ asymptotic value of the dilaton is time-independent and its behavior at $v=0$ is regular.
Note also that this turns out to be the selected value of $\kappa$ in Choptuik's critical collapse of a massless scalar field~\cite{Gundlach:1999cu}.

Let us first analyze the unsourced Maxwell equation, Eq.~\eqref{MaxwellEOM} with $J^\nu=0$. The only non-trivial CSS equations one finds arise from the $v$- and $r$-components,
\bea
& (B \overline{Q})' - 2aB\overline{Q}\phi' =0\,,\\
& (B \overline{Q})' + B\overline{Q} \left( \frac{1}{\zeta} - 2a\phi' \right) =0\,.
\eea
The only way these two conditions can be consistent is if $\overline{Q}(\zeta)=0$, i.e., the electromagnetic field must vanish.

In the absence of an electromagnetic field the remaining CSS equations of motion reduce to the following set of equations (which are not all independent):
\begin{flalign}
& \frac{B'^2}{2B^2} - \frac{B''}{B} - 2\phi'^2  = 0\,, \label{eq:CSSvacuum1}\\
& \zeta A'' - 4\phi'^2 + \left( \zeta A' \frac{B'}{B} - 2\frac{B''}{B} \right) + \frac{B'^2}{B^2} = 0\,, \label{eq:CSSvacuum2}\\
& 1 - B' + \zeta B'' +\zeta B A' - \frac{\zeta^2}{2} A' B' - A \left(B - \zeta B' + \frac{\zeta^2}{2} B'' \right) = 0\,, \\
& \zeta^2 A A' \frac{B'}{B} - 2 \frac{B''}{B} + \frac{B'^2}{B^2}  + \zeta^2 A A'' - 2 A' - 4\phi'^2 = 0\,, \\
& (\zeta A - 2) \left( \phi''  + \frac{B'}{B} \phi' \right) + \zeta A' \phi' =0\,. \label{eq:CSSvacuumDil}
\end{flalign}
These equations can be solved exactly. A linear combination of~\eqref{eq:CSSvacuum1} and~\eqref{eq:CSSvacuum2} reveals that
\be
\left(B A'\right)'=0\,.
\ee
The only solution that is consistent with the other equations is $A(\zeta)=1+2\nu=const$. Consistency of the other option, $B \propto A'^{-1}$, would require the vanishing of $B$, which does not yield a sensible line element. Plugging a constant $A$ in the dilaton equation~\eqref{eq:CSSvacuumDil} immediately gives
\be
\phi' = \frac{-\sigma}{B}\,,
\label{eq:phiprime}
\ee
where $\sigma$ is a constant. Finally, using the remaining equations we obtain
\be
B(\zeta) = 1 - 2\nu\zeta + \left(\nu^2-\sigma^2\right)\zeta^2\,,
\ee
where $\nu$ is the previous integration constant. The other constant of integration has been fixed so that $r$ corresponds asymptotically to the areal radius, which amounts to requiring $B(\zeta=0)=1$.

To complete the solution we must integrate Eq.~\eqref{eq:phiprime} to obtain $\phi(\zeta)$. Before doing so, it is convenient to make the coordinate transformation $r \to r - (\sigma-\nu)v$, which implies $\zeta \to \frac{\zeta}{1 - (\sigma-\nu)\zeta}$. The advantage of this transformation is that it reduces the function $B$ to a linear form, thus simplifying the task of integrating~\eqref{eq:phiprime}. The final result is then
\begin{subequations}
\label{eq:Roberts}
\bea
ds^2 &=& -(1+2\sigma)dv^2 + 2dvdr + r^2 \left( 1-\frac{2\sigma v}{r} \right) d\Omega^2\,, \\
e^{2\Phi} &=& 1-\frac{2\sigma v}{r}\,.
\eea
\end{subequations}
In general, a further integration constant would appear in the latter equation, but it can be absorbed by a trivial shift of the dilaton.
This is recognized as Roberts' scale-invariant solution~\cite{Roberts:1989sk} to the Einstein-dilaton system. Thus, we have shown that the most general source-free, spherically symmetric, continuously self-similar solution (with $\kappa=0$) of EMD theory necessarily has vanishing electromagnetic field and is none other than Roberts' spacetime.

\section{CSS solutions in the heterotic theory supported by null dust}
\label{sec:CSSheterotic}

Now we investigate continuously self-similar solutions of EMD theory supported by charged null dust.
In this section we fix the dilaton coupling to the heterotic string value $a=1$, but this restriction will be lifted in sections~\ref{sec:genericCSS} and~\ref{sec:CSSsourcedSols}.

\subsection{The `linear' solutions} \label{sub_sec:linear_solutions}

A simple family of self-similar solutions sourced by a charged null fluid is obtained by taking all the functions $A$, $B$, $Q$ and $e^{2\phi}$ in Eqs.~\eqref{eq:ansatz_metric} and~\eqref{eq:ansatz_matter} to be linear in the variable $\zeta$ (recall we are fixing $\kappa=0$),
\begin{subequations}
\bea \label{eq:linearsol_metric}
ds^2 &=& -\left(1+ 2\sigma - \frac{q^2v}{\sigma r} \right) dv^2 + 2 dv dr + r^2 \left(1-\frac{2\sigma v}{r}\right) d\Omega^2\,,\\ \label{eq:linearsol_Maxwell}
F &=& -\frac{q v}{r^2} dv\wedge dr \,,\\
 e^{2\Phi} &=& 1-\frac{2\sigma v}{r}\,.\label{eq:linearsol_dilaton}
\eea
\end{subequations}
These solutions are defined by two constant parameters, namely $\sigma$ and $q$. Alternatively, they can also be recovered by applying Vaidya's procedure to the time-dependent vacuum spacetimes derived in~\cite{Aniceto:2017gtx} for $a=1$ [see Eqs. (4.1-2)]. Those solutions have constant electric charge $Q$, while the mass parameter $M(v)$ is constrained to be inversely proportional to the dilaton charge $D(v)$. However, promoting the electric charge to be a linear function of advanced time, $Q(v)=qv$, allows for the mass and dilaton charge to also grow linearly in time, hence producing a self-similar spacetime. By doing so, one introduces source terms in the equations of motion:
\begin{subequations}
\bea
J^\nu &=& - \frac{q}{4\pi r^2 (1-2\sigma v/r)} \delta^\nu_r\,, \label{eq:Jnu}\\
T_{\mu\nu}^{\rm (fluid)} &=& \mu \ell_\mu \ell_\nu\,, 
\qquad {\rm where} \quad \mu= \frac{q^2}{\sigma r^2}
\qquad {\rm and} \quad \ell_\mu = (-1,0,0,0)\,. \label{eq:Tmunu}
\eea
\end{subequations}
The source supporting these solutions thus corresponds to charged null dust and it satisfies the null, weak, strong and dominant energy conditions as long as $\sigma>0$. This guarantees that $\mu>0$, which is sufficient to comply with the energy conditions for a stress-energy tensor corresponding to null dust. However, the total stress-energy tensor $T_{\mu\nu}$ also includes contributions from the dilaton and Maxwell fields, so the analysis of the energy conditions requires a bit more effort. Notwithstanding, it is shown in section~\ref{sec:EnergyConds} and appendix~\ref{App1} that all energy conditions are satisfied for the spacetime defined by the line element~\eqref{eq:linearsol_metric}, as long as $\sigma>0$~\footnote{Although the solutions presented in this section are not quite the same as those obtained in section~\ref{sec:CSSsourcedSols}, the metric is manifestly the same when $a=1$, and this is all that matters to assess the energy conditions for the total stress-energy tensor.}.

The source terms~\eqref{eq:Jnu} and~\eqref{eq:Tmunu} display the correct form to leave the equations of motion invariant under rescalings $(v,r)\to (kv , kr)$, as should be expected for a self-similar solution.

\subsection{Critical exponent from the `linear' solutions}

It is possible to employ these CSS solutions sourced by null dust in order to investigate critical behavior in EMD theory. The procedure parallels Brady's approach~\cite{Brady:1994xfa}, which was developed specifically for the Einstein-massless scalar system based on the previously known Roberts' class of CSS spacetimes~\cite{Roberts:1989sk}.
The idea is to consider an initially flat space on which we turn on some influx of null matter at an advanced time $v=0$. This scenario can be described in the context of EMD theory (with dilaton coupling $a=1$) by the continuous matching of Minkowski spacetime to the `linear' solutions presented above. The matching is done along the null hypersurface $v=0$.

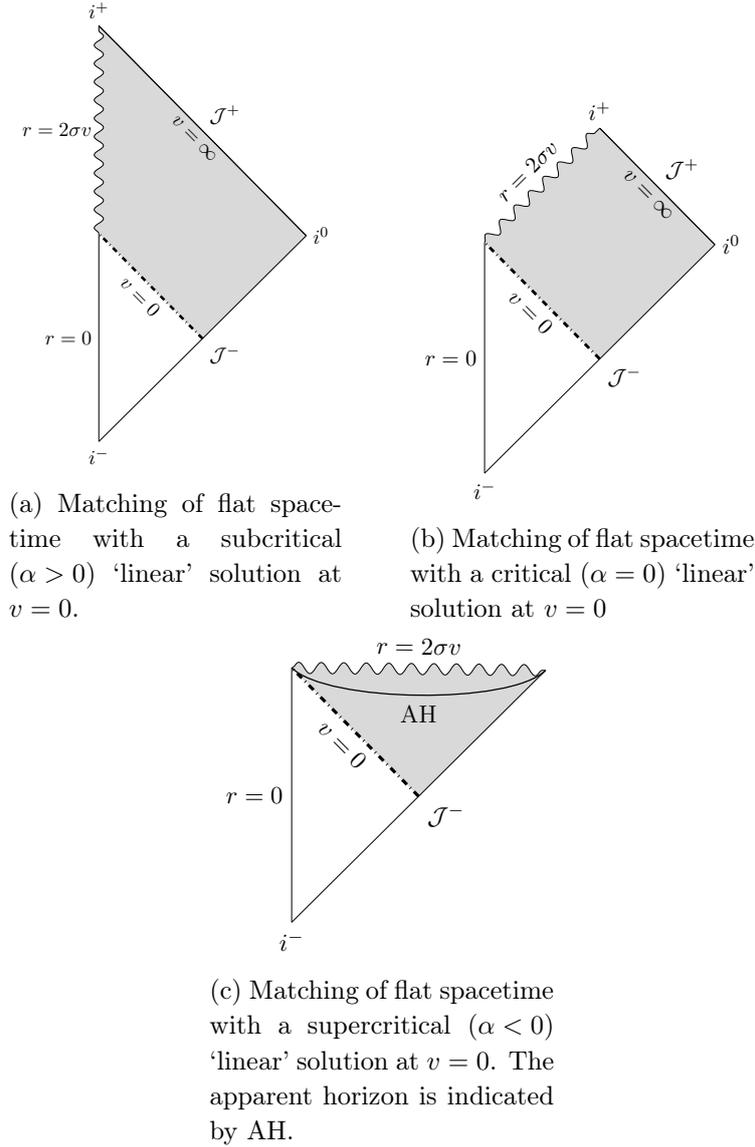
\begin{figure}[!t]
	\centering	
	\begin{subfigure}[b]{0.29\textwidth}
	\centering
	\resizebox{\linewidth}{!}{
		\begin{tikzpicture}

		\node (I)    at (-1,0)   {};
		
		\path
		(I) + (-90:1) coordinate[label=-90:$i^-$] (II);
		
		\path
		(II) + (45:2.8) coordinate (IIright)
		+ (90:4) coordinate (IItop)
		+  (45:5.6) coordinate[label=0:$i^0$] (IIIright)
		+ (90:8) coordinate[label=90:$i^+$] (IIItop);

		\begin{scope}[decoration=snake]
		\draw[draw=none,fill=gray!30] 	(IItop) decorate{--	(IIItop) }
		-- (IIIright)
		-- (IIright)
		-- (IItop);
		\end{scope}

		\draw[decorate,decoration=snake] (IItop) --
		node[midway, left] {$r=2\sigma v$}	(IIItop) ;
		
		\draw
		(II) --
		node[midway, left]    {$r=0$}
		(IItop);
		
		\draw
		(II) -- 
		node[midway, below right]    {$\cal{J}^-$}
		(IIIright);			
		
		\draw[dashdotted, line width=0.5mm]
		(IIright) to
		node[midway, below, sloped]    {$v=0$}
		(IItop);

		\draw (IIIright) -- 
		node[midway, above right] {$\cal{J}^+$}
		(IIItop);
		
		\draw (IIIright) -- 
		node[midway, below, sloped] {$v= \infty$}
		(IIItop);

		\end{tikzpicture}
	}
	\caption{Matching of flat spacetime with a subcritical $\left(\alpha >0\right)$ `linear' solution at $v= 0$.}
	\label{fig_matching:subcritical} 	
		\end{subfigure} \qquad 	
	\begin{subfigure}[b]{0.3\textwidth}
		\centering
		\resizebox{\linewidth}{!}{
			\begin{tikzpicture}

			\node (I)    at (-1,0)   {};
			
			\path
			(I) + (-90:1) coordinate[label=-90:$i^-$] (II);
			
			\path
			(II) + (45:2.8) coordinate (IIright)
			+ (90:4) coordinate (IItop)
			+  (45:5.6) coordinate[label=0:$i^0$] (IIIright);

			\path 
		 (IItop) + (45:2.8) coordinate[label=90:$i^+$] (IIItop);

			\begin{scope}[decoration=snake]
			\draw[draw=none,fill=gray!30] 	(IItop) decorate{--	(IIItop) }
			-- (IIIright)
			-- (IIright)
			-- (IItop);
			\end{scope}
			
			\draw[decorate,decoration=snake] (IItop) --
			node[midway, sloped, above=0.1cm ] {$r=2\sigma v$}
			(IIItop);

			\draw
			(II) --
			node[midway, left]    {$r=0$}
			(IItop);
			
			\draw
			(II) -- 
			node[midway, below right]    {$\cal{J}^-$}
			(IIIright);			
			
			\draw[dashdotted, line width=0.5mm]
			(IIright) to
			node[midway, below, sloped]    {$v=0$}
			(IItop);

			\draw (IIIright) -- 
			node[midway, above right] {$\cal{J}^+$}
			(IIItop);
		
			\draw (IIIright) -- 
			node[midway, below, sloped] {$v= \infty$}
			(IIItop);
			
			\end{tikzpicture}
		}
		\caption{Matching of flat spacetime with a critical $\left(\alpha =0\right)$ `linear' solution at $v= 0$}
		\label{fig_matching:critical} 	
	\end{subfigure}  \qquad 
	\begin{subfigure}[b]{0.3\textwidth}
	\centering
	\resizebox{\linewidth}{!}{
		\begin{tikzpicture}

		\node (I)    at (-1,0)   {};
		
		\path
		(I) + (-90:1) coordinate[label=-90:$i^-$] (II);
		
		\path
		(II) + (45:2.8) coordinate (IIright)
		+ (90:4) coordinate (IItop)
		+  (45:5.6) coordinate (IIIright);

		\begin{scope}[decoration=snake]
		\draw[draw=none,fill=gray!30] 	(IItop) decorate{--	(IIIright) }
		-- (IIright)
		-- (IItop);
		\end{scope}

		\draw[decorate,decoration=snake] (IItop) --
		node[midway, sloped, above=0.1cm ] {$r=2\sigma v$}
		(IIIright);

		\draw
		(II) --
		node[midway, left]    {$r=0$}
		(IItop);
		
		\draw
		(II) -- 
		node[midway, below right]    {$\cal{J}^-$}
		(IIIright);			
		
		\draw[dashdotted, line width=0.5mm]
		(IIright) to
		node[midway, below, sloped]    {$v=0$}
		(IItop);					
		
		\draw[line width=0.2mm]
		(IItop) to[out=-45, in=225, looseness=0.5] 
		node [below, midway, inner sep=1.5mm] {AH}  
		(IIIright);								
		
		\end{tikzpicture}
	}
	\caption{Matching of flat spacetime with a supercritical $\left(\alpha < 0\right)$ `linear' solution at $v= 0$. The apparent horizon is indicated by AH.}
	\label{fig_matching:supercritical} 	
\end{subfigure} 
	\caption[Penrose diagrams]{Carter-Penrose diagrams corresponding to the three qualitatively different types of collapse obtained by matching flat spacetime to the `linear' CSS solutions sourced by null dust along the null hypersurface $v=0$.} 
	\label{fig:Penrose}
\end{figure}

In the present case, the `linear' solutions we adopt are characterized by two free parameters. This should be contrasted with Ref.~\cite{Brady:1994xfa}, where the self-similar solutions used to construct the global spacetime featured only one parameter. Within our two-dimensional parameter space certain regions will describe the formation of a black hole\footnote{As usually occurs for spacetimes enjoying homothetic symmetry, the mass of the black hole formed grows linearly in time.} while other regions never develop an apparent horizon.
Note however that in the latter case we still form a curvature singularity (see below), which is therefore naked.\footnote{This is another point in which our study departs from~\cite{Brady:1994xfa}, where in the subcritical regime of parameters a second matching to Minkowski space is done once the (radially) inflowing matter goes through the origin and starts dispersing, leaving behind flat space again. Instead, Fig.~\ref{fig:Penrose} (a) reproduces the situation considered in Ref.~\cite{Roberts:1989sk}, but now in the context of EMD theory.} At the frontier between these two regions lies the critical set of parameters for which a null singularity is produced, corresponding to the threshold of black hole formation. This general picture is illustrated in Fig.~\ref{fig:Penrose}.

Clearly, we then need to determine the causal structure of the spacetime defined by~\eqref{eq:linearsol_metric}, to which we now turn our attention.

The line element~\eqref{eq:linearsol_metric} features curvature singularities at $r=2\sigma v$, as well as at $r=0$. Since we are interested in matching these solutions onto flat space along the null hypersurface $v=0$, we will consider $v\geq0$. Restricting to positive values of $\sigma$, the only singular surface of physical relevance is therefore
\be 
r=r_s(v) \equiv 2\sigma v
\qquad \Longleftrightarrow \qquad
\zeta=\zeta_s \equiv (2\sigma)^{-1}\,.
\ee
The character of the singular surface $S(v,r) \equiv r-r_s(v)$ is dictated by the sign of the norm of the vector perpendicular to $S$ and is found to be
\be
\left\{
\begin{array}{lc}
{\rm spacelike \; if}  &  q^2 > 2\sigma^2 (1-2\sigma)\,,  \\
{\rm null \; if}           &  q^2 = 2\sigma^2 (1-2\sigma)\,,  \\
{\rm timelike \; if}    &  q^2 < 2\sigma^2 (1-2\sigma)\,.
\end{array}
\right. \nn
\ee

On the other hand, the location of an apparent horizon is determined by a vanishing expansion of radial outgoing null geodesics. For the class of metrics we are considering, this entails the following condition~\cite{Aniceto:2017gtx}:
\be
A\left(2B+r\partial_r B\right) +2r\partial_v B = 0\,.
\label{eq:conditionAH}
\ee
Plugging in the metric components read from~\eqref{eq:linearsol_metric}, it is easy to show that an apparent horizon only exists when
\be
q^2 > 2\sigma^2 (1-2\sigma)\,,
\ee
i.e., when the singularity is spacelike.
Otherwise it occurs `inside' the singular surface and is therefore unphysical.
Expressed in terms of the scaling variable, the apparent horizon is located at
\be
\zeta_{ah} = \frac{1}{2\sigma}\left\{1+\frac{\sigma^2(1+2\sigma)}{q^2} - \sqrt{\left( 1+\frac{\sigma^2(1+2\sigma)}{q^2} \right)^2 - \frac{4\sigma^2}{q^2}}\right\}\,.
\ee

One can now take these solutions and continuously match them onto flat space along $v=0$, closely following Ref.~\cite{Brady:1994xfa}. One necessarily gets a surface distribution of scalar charge along the junction. Depending on the parameters $\sigma$ and $q$ these spacetimes describe the formation of a naked singularity (if $q^2 < 2\sigma^2 (1-2\sigma)$), or the formation of an apparent horizon (if $q^2 > 2\sigma^2 (1-2\sigma)$). The critical case, lying at the threshold $q^2 = 2\sigma^2 (1-2\sigma)$, corresponds to the formation of a null singularity.

The relevant quantity for us is the mass enclosed in the apparent horizon. In particular, we are interested in computing this as a function of the {\it criticality parameter}, which we define as
\be
\alpha \equiv 2\sigma^2 (1-2\sigma) - q^2\,.
\label{eq:defalpha}
\ee
The critical solution then corresponds to $\alpha=0$. Note that in this case (and only in this case) the apparent horizon coincides with the singularity,
\be
\left.\zeta_{ah}\right|_{\alpha=0} = \frac{1}{2\sigma} = \zeta_s\,.
\ee
The quantity $-\alpha$ plays the role of the parameter $p$ described in the Introduction.

Evaluating the local mass function at the apparent horizon and expanding around the critical point we find
\be
M_{ah}(v) = \frac{v}{2} (-\alpha)^{1/2} \sqrt{\frac{2\left[1+O(-\alpha)\right]}{1+2\sigma}}\,.
\ee
This blows up linearly in null time $v$ but one can obtain a finite quantity by normalizing with respect to the total mass (inferred at $r=\infty$), which also grows linearly.
Therefore, we read off the critical exponent as being $\gamma=1/2$, the same value as in the Einstein-dilaton case~\cite{Brady:1994xfa}.
Hence, even with the addition of an electric field and of sources for the field equations we still get the same critical exponent as in the neutral source-free case.

However, as we will show below, this is not a generic statement, even in the context of Einstein-Maxwell-dilaton theory. In fact, it is specific to the choice of dilaton coupling $a=1$.

\section{Critical exponent from generic CSS spacetimes}
\label{sec:genericCSS}

Now we will generalize the previous section by allowing the dilaton coupling $a$ to be arbitrary.

The only information relevant to determine the (mass) critical exponent is the metric. The Maxwell and dilaton fields play no role here; they are important only for the equations of motion and consequently only to determine the kind of sources supporting the solutions.
So take
\be \label{eq:sec_crit_exp_metric}
ds^2 = -A(\zeta) dv^2 + 2 W(\zeta)dv dr + r^2 B(\zeta) d\Omega^2\,, \qquad \zeta\equiv\frac{v}{r}\,.
\ee
This line element is continuously self-similar, with homothetic vector field $\xi=v\frac{\partial}{\partial v}+r\frac{\partial}{\partial r}$.
In order to be able to probe the critical regime ---i.e., form black holes with arbitrarily small masses--- it is crucial that the ratio $f(\zeta)\equiv A(\zeta)/W(\zeta)$ is analytic at the singularity, as will become evident below.

Such a spacetime features curvature singularities at $r=0$ and at $\zeta_s$ such that
\be
B(\zeta_s) = 0\,,
\ee
since the curvature scalar is proportional to $r^{-2} B(\zeta)^{-2} W(\zeta)^{-3}$. There might also exist other singular points $\zeta_*$ ---for example, at $W(\zeta_*)=0$. We assume that they all satisfy $\zeta_*\geq \zeta_s$ so any other possible singular points are not part of the spacetime.

The character of the singular surface $\zeta=\zeta_s$ is determined by the sign of the norm of the normal vector,
\be
\left.(\nabla B)^2\right|_{B=0} = \frac{B'(\zeta_s)^2 \zeta_s}{W(\zeta_s)^2 r_s^2} \big[ \zeta_s A(\zeta_s) - 2 W(\zeta_s) \big]\,,
\ee
from which we conclude that the singularity is
\be
\left\{
\begin{array}{lc}
{\rm spacelike \; if}  &  \zeta_s \left[ \zeta_s A(\zeta_s) - 2 W(\zeta_s) \right] < 0\,,  \\
{\rm null \; if}           &  \zeta_s \left[ \zeta_s A(\zeta_s) - 2 W(\zeta_s) \right] = 0\,,  \\
{\rm timelike \; if}    &  \zeta_s \left[ \zeta_s A(\zeta_s) - 2 W(\zeta_s) \right] > 0\,.
\end{array}
\right. \nn
\ee
The requirement of the singularity being spacelike is actually tied to the existence of an apparent horizon. It is then natural to define a critical parameter by
\be
\alpha \equiv \zeta_s \, f(\zeta_s) - 2\,.
\ee

In case it exists, an apparent horizon is located at the hypersurface $\zeta=\zeta_{ah}$ where the expansion of outgoing null geodesics vanishes. For the class of metrics~\eqref{eq:sec_crit_exp_metric}, it can be shown that this condition is equivalent to
\be
A(\zeta_{ah}) \Big[2B(\zeta_{ah})-\zeta_{ah} B'(\zeta_{ah})\Big] + 2 W(\zeta_{ah}) B'(\zeta_{ah}) = 0\,,
\label{eq:ApHor}
\ee
or expressed in terms of $f$,
\be
\left[ \zeta_{ah} \, f(\zeta_{ah}) - 2 \right] B'(\zeta_{ah}) = 2 f(\zeta_{ah}) B(\zeta_{ah})\,.
\label{eq:ApHor2}
\ee
Clearly, when the singular surface is null the locus $\zeta_{ah}=\zeta_s$ is a solution to the above equation, since in that case both sides of this equation vanish.

The mass enclosed in the apparent horizon can be computed by
\be
M(v,r_{ah}) = \frac{R(v,r_{ah})}{2} \left( 1-(\nabla R)^2 \right)_{r=r_{ah}} 
  = \frac{v \sqrt{B(\zeta_{ah})}}{2\zeta_{ah}}\,,
\ee
where $R(v,r)^2= r^2 B(\zeta)$. For late advanced times, this diverges linearly with $v$. Dividing by the mass evaluated at $r\to\infty$ as in~\cite{Brady:1994xfa}, one obtains a finite quantity.
A large-$r$ expansion (equivalently, small-$\zeta$ expansion) yields
\be
M(v,r) = \frac{v}{2\zeta} \sqrt{B_0} \left[ \mathcal{M}^{(-1)} + \mathcal{M}^{(0)}\,\zeta + O(\zeta^2) \right]\,,
\ee
where
\begin{flalign}
&\mathcal{M}^{(-1)} \equiv 1-\frac{A_0 B_0}{W_0^2}-\frac{B'_0}{W_0}\,, \\
&\mathcal{M}^{(0)} \equiv \frac{B_0 \left[2 A_0 W'_0-W_0 A'_0\right]}{W_0^3} + \frac{B'_0 \left[2 A_0 (1-B_0)+2B_0W'_0+W_0^2\right]}{2 B_0 W_0^2}-\frac{B''_0}{W_0}\,,
\end{flalign}
and we have defined $A_0=A(0), B_0=B(0), W_0=W(0), \dots,$ to avoid additional cluttering of the equations.
This shows that a finite total mass (at any given advanced time $v$) requires
\be
W_0^2 - A_0 B_0 - W_0 B'_0 =0\,,
\label{eq:FiniteMass}
\ee
and in that case
\be
M(v,r\to\infty) = \frac{v}{2} \sqrt{B_0}\, \mathcal{M}^{(0)}\,.
\ee
Hence, 
\be
\frac{M(v,r_{ah})}{M(v,r\to\infty)} = \frac{1}{\sqrt{B_0}\, \mathcal{M}^{(0)}} \sqrt{\frac{B(\zeta_{ah})}{\zeta_{ah}^2}}\,.
\label{eq:MassRatio}
\ee

Now, in order to infer the critical exponent we need to be able to express this ratio in terms of the parameter $\alpha$ measuring the departure from criticality. It turns out this can be done ---under some assumptions--- for a broad class of solutions without specifying the exact form of the metric functions $A$ or $W$; the only crucial input enters through function $B$ that controls the size of spheres of constant $v$ and $r$. To be concrete, the assumptions are:
\begin{itemize}
\item The metric function $B$ behaves as $B(\zeta) = \left(1-\frac{\zeta}{\zeta_s}\right)^{2\delta}$, at least near $\zeta_s$. This is in fact what one naturally gets by applying Vaidya's procedure to the static GMGHS solution to obtain sourced spacetimes, as we shall do in the following section. In that case $\delta=\frac{a^2}{1+a^2}$.
\item The function $f(\zeta)=A(\zeta)/W(\zeta)$ is analytic at the singularity, $\zeta=\zeta_s$.
\end{itemize}

Under these conditions, we can determine $\zeta_{ah}$ perturbatively in $\alpha$ by performing a series expansion of Eq.~\eqref{eq:ApHor2} around $\zeta_s$. The result of this calculation is
\be
\alpha + \left(\zeta_{ah}-\zeta_s\right) \left\{ \zeta_s\, f'(\zeta_s) - \frac{f(\zeta_s)}{a^2} \right\} + O\left(\zeta_{ah}-\zeta_s\right)^2 = 0\,.
\ee
Inverting this for $\zeta_{ah}$ yields
\be
\zeta_{ah} = \zeta_s + \frac{\alpha}{\left[ \frac{f(\zeta_s)}{a^2} - \zeta_s\, f'(\zeta_s) \right]_{\alpha=0}} + O(\alpha^2)\,,
\ee
which is valid as long as $C \equiv \left[\zeta_s \frac{f(\zeta_s)}{a^2} - \zeta_s^2 f'(\zeta_s) \right]_{\alpha=0} \neq0$.
We can now plug this back in Eq.~\eqref{eq:MassRatio} to obtain
\be
\frac{M(v,r_{ah})}{M(v,r\to\infty)} \propto \sqrt{B(\zeta_{ah})} = \left( 1-\frac{\zeta_{ah}}{\zeta_s} \right)^\delta = C^{-\delta} (-\alpha)^\delta + O(\alpha^{1+\delta}) \,,
\ee
indicating that the (mass) critical exponent is precisely given by $\delta$.

\section{CSS solutions supported by null fluids with arbitrary dilaton coupling}
\label{sec:CSSsourcedSols}

In section~\ref{sub_sec:linear_solutions} we presented CSS solutions for the case in which the dilaton coupling value is $a=1$. Nonetheless, CSS solutions with {\em arbitrary} dilaton coupling may also be found by generalizing the 2-parameter family of static GMGHS solutions~\cite{Gibbons:1987ps,Garfinkle:1990qj},
\begin{subequations}
\bea 
ds^2 &=& -\left(1-\frac{r_+}{r}\right) \left(1-\frac{r_-}{r}\right)^{\frac{1-a^2}{1+a^2}} dv^2 + 2dv dr + r^2 \left(1-\frac{r_-}{r}\right)^{\frac{2a^2}{1+a^2}} d\Omega^2\,,\\
F &=& - \sqrt{\frac{r_+ r_-}{1+a^2}} \frac{1}{r^2} dv \wedge dr \,,\\
e^{2 a \Phi} &=& \left(1 - \frac{r_-}{r}\right)^\frac{2a^2}{1+a^{2}}  \,.
\eea
\end{subequations}

By promoting the constant parameters $r_+$ and $r_-$ to be functions of $v$ and $r$ one obtains sourced solutions, i.e, with additional sources $T_{\mu\nu}^\text{(fluid)}, J^\nu, \Sigma$ in the field equations~\eqref{eq:fieldeqs}. It turns out that in order for the extra contribution $T_{\mu\nu}^\text{(fluid)}$ to the stress-energy tensor to be a Husain-type null fluid, $r_-$ must be a function linear in the radial coordinate, $r_-(v,r)=r_-(v)+c(v)r$. Without loss of generality we can set $c(v)=0$, since its effect just amounts to rescaling the radial coordinate. Restricting to self-similar spacetimes then selects $r_-(v)=m v$, with $m$ being a constant.
This procedure does not fix the form of the metric function $A$, but as we saw above we want $A/W$ to be analytic (at the very least it must be finite) at the singular point $\zeta=\zeta_s=m^{-1}$ in order to address critical collapse. For $a\neq1$, the factor $\left(1-\frac{r_-}{r}\right)^{\frac{1-a^2}{1+a^2}}$ appearing in $A$ complicates matters, but this obstacle can be circumvented by promoting $r_+$ to a function of the coordinates $v$ and $r$ according to
\be
r_+(v,r) = r \left[ 1- \frac{f(\zeta)B(\zeta)}{1-m\zeta} \right]\,,
\ee
with $f$ an analytic function at $\zeta=\zeta_s$.
This yields a CSS metric of the form~\eqref{eq:sec_crit_exp_metric} with
\be
\label{eq:a_neq_1_metric}
A(\zeta)= f(\zeta) \,, \qquad
W(\zeta)=1 \,, \qquad
B(\zeta)= \left(1- m \zeta\right)^\frac{2 a^2}{1+a^2} \,.
\ee

As it stands, the metric function $A$ is still unfixed.
The simplest choice, consistent with condition~\eqref{eq:FiniteMass}, is to take $f(\zeta)$ to be linear in $\zeta,$
\be
f(\zeta)= 1+\frac{2a^2m}{1+a^2}-p\zeta\,,
\ee
thus yielding a two-parameter family of solutions, here denoted by $m$ and $p$. Henceforth, we shall assume that $f'(\zeta) =-p < 0$.

For the Maxwell and dilaton field we then obtain
\be
F = - \sqrt{\frac{m}{1+a^2}} \left[ \frac{1}{\zeta}- \frac{f(\zeta)B(\zeta)}{\zeta(1-m\zeta)} \right]^{1/2} \frac{v}{r^2} dv \wedge dr \,, 
\qquad
e^{2 a \Phi} = \left(1 - m \zeta\right)^\frac{2a^2}{1+a^{2}}  \,. \label{eq:a_neq_1_dilaton}
\ee
Additionally, the non-null components of the sources for the dilaton and Maxwell fields are given by
\begin{subequations}
\bea
\Sigma  &=& \frac{a m \zeta  \left[  \left(1-m \zeta\right)^\frac{2}{1+a^2}    -\left(\frac{1-a^2}{1+a^2}\right)2m - \left(1-\frac{2m\zeta}{1+a^2} \right)f(\zeta) - p\zeta \left(1-m \zeta\right) \right] }{4 \pi r^2  \left(1+a^2\right) \left(1-m \zeta\right)^2}  \,, \\
J^v &=&  \frac{m \left[ \left(1-\frac{2m\zeta}{1+a^2} \right) f(\zeta) + p\zeta \left(1-m \zeta \right)  - \left(1- m\zeta\right)^\frac{2}{1+a^2}  \right]}{8 \pi r^2 \left(1+a^2\right) \sqrt{\frac{m}{1+a^2}} \left(1-m \zeta \right)\left[ \frac{1}{\zeta}- \frac{f(\zeta)B(\zeta)}{\zeta(1-m\zeta)} \right]^{1/2} }  \,,  \\
J^r &=& -\frac{m \left[ \left(1-\frac{2a^2m\zeta}{1+a^2} \right) f(\zeta) - p\zeta \left(1-m \zeta \right) - \left(1- m\zeta\right)^\frac{2}{1+a^2}  \right]}{8 \pi r^2 \left(1+a^2\right) \zeta \left(1-m \zeta \right)\sqrt{\frac{m}{1+a^2}} \left[ \frac{1}{\zeta}- \frac{f(\zeta)B(\zeta)}{\zeta(1-m\zeta)} \right]^{1/2} } \,.
\eea
\end{subequations}
Finally, the fluid stress-energy tensor takes the form of a Husain null fluid~\cite{Husain:1995bf}. To describe this fluid we begin by introducing two future-pointing null vectors,
\be
\ell_\mu = - \delta^v_\mu \,, \qquad n_\mu = \frac{1}{2} g_{vv} \delta^v_\mu + \delta^r_\mu \,.
\ee
The fluid stress-energy tensor is then given by
\be
T_{\mu\nu}^{\rm (fluid)} = \mu\,  \ell_\mu \ell_\nu + 2 \left(\rho + P\right)\,  \ell_{(\mu}  n_{\nu)}  + P g_{\mu \nu } \,,
\ee
where we have defined 
\begin{flalign}
\label{eq:aneq1_mu_value}
& \mu =   \frac{p}{8 \pi r^2}	 \,,  \\
& \rho =  \frac{\frac{2a^2 m}{1+a^2} \left(1- \frac{2m\zeta}{1+a^2} \right) + \left(1 - \frac{m \zeta}{1+a^2} \right) \left[\left(1-m\zeta\right)^\frac{2}{1+a^2}  -p \zeta \left(1-m\zeta\right) - \left(1- \frac{2m\zeta}{1+a^2}\right)f(\zeta) \right]}{8 \pi r^2 \left(1-m\zeta\right)^2 }	      \label{eq:aneq1_rho_value}	  \,, \\
& P =  \frac{m \zeta \left[ a^2 p \zeta+ f(\zeta ) - \left(1-m \zeta\right)^\frac{1-a^2}{1+a^2}  \right]}{8 \pi r^2 \left(1+a^2\right) \left(1-m \zeta\right)} \,.
\end{flalign}
This stress-energy tensor describes a  null fluid with energy flux $\mu$ moving with four velocity $\ell^\mu$. Since $\ell_\mu \ell^\mu = n_\mu n^\mu =0$ and $\ell_\mu n^\mu = -1$ we have that the stress-energy tensor supports an energy flux only along the null vector $n$.  The energy conditions for this type of tensor have been carefully spelled out in Ref.~\cite{Creelman:2016laj}, for instance. For $\mu\neq0$ ---corresponding to the case of interest to us--- this stress-energy tensor is of type~II according to the terminology of~\cite{Hawking:1973uf}. In any case, in order to determine whether energy conditions are violated or not one needs to consider the {\em total} stress-energy tensor. This will be tackled in section~\ref{sec:EnergyConds} below.

Before moving on to the analysis of the causal structure of the spacetime and the resulting critical behavior, we should note that the solutions presented in this section do {\em not} reduce to the previous solutions displayed in section~\ref{sec:CSSheterotic} when taking $a=1$. In fact, setting $a=1$ in the above expressions returns
\be
\rho=P=\frac{m^2\zeta}{16\pi r^2 (1-m\zeta)}\neq0\,,
\ee
which does not correspond to null dust as in section~\ref{sec:CSSheterotic}. Moreover, the electric field is not directly proportional to $\frac{v}{r^2}$, which was the case in Eq.~\eqref{eq:linearsol_Maxwell}.

\subsection{Causal structure of the spacetime and critical exponent} 

As in section~\ref{sec:CSSheterotic}, we now want to investigate critical collapse by matching these CSS solutions sourced by null fluids onto flat space along the null hypersurface $v=0$. In order to determine what choices of parameters lead to black hole formation and those for which a horizon is never produced, we first need to understand the causal structure of the spacetime, namely the character of the singularity and the location of the apparent horizon (if it exists).

The character of an hypersurface is conformally invariant. 
To figure out the character of the singularity one must compute the norm of the normal vector to the surface defined by $B(\zeta)=0$. However, that expression turns out to be proportional to $(1-m\zeta_s)^{2\frac{a^2-1}{a^2+1}}$ and so for $a\neq1$ it will either vanish or blow up. It is then preferable to work in a different conformal frame where this overall factor is absent. Equivalently, we can instead compute the norm of the normal vector to the hypersurface $B(\zeta)^{\frac{1+a^2}{2a^2}}=0$. The character of the hypersurface is determined by the sign of
\be
\label{eq:arbitrary_a_character_singularity}
\lim_{\zeta \to \frac{1}{m}} \left[ \nabla \left(B(\zeta)^{\frac{1+a^2}{2a^2}} \right)\right]^2  =
\frac{1}{r_s^2} \left[ 1-\frac{2m}{1+a^2} - \frac{p}{m} \right]\,.
\ee
The singular surface is then
\be
\left\{
\begin{array}{lccc}
	{\rm spacelike \; if}  &  p > m\left(1-\frac{2m}{1+a^2}\right)\,,  \\
	{\rm null \; if}           &  p = m\left(1-\frac{2m}{1+a^2}\right) \,, \\
 	{\rm timelike \; if}    &  p < m\left(1-\frac{2m}{1+a^2}\right)\,. 
\end{array}
\right. \nn
\ee
The case $a=1$ matches precisely what we obtained before. This becomes evident after taking $p= q^2/\sigma$ and $m=2\sigma$.

The location of the apparent horizon is again determined by Eq.~\eqref{eq:ApHor}, which for these solutions simply reduces to
\be
\left(1+a^2\right)\left(1-p \zeta \right) \left(1+a^2- m \zeta\right) -2 a^2 m^2 \zeta= 0\,. \label{eq:cond_AH_arbitrary_a}
\ee
This quadratic equation can be solved for $\zeta_{ah}$, yielding explicitly
\be
\zeta_{ah} = \frac{1}{m} + \frac{m+\frac{2a^2m^2}{1+a^2}  + \left(a^2-1\right)p - \sqrt{\left[m+\frac{2a^2m^2}{1+a^2}  + \left(a^2-1\right)p\right]^2 +4a^2 p \left[p -m\left(1-\frac{2m}{1+a^2}\right)\right]}}{2mp}\,.
\ee
When $p/m=1-2m/(1+a^2)$ we get $\zeta_{ah} = m^{-1} = \zeta_s$. In this case (and only in this case) the apparent horizon is null and coincides with the singular surface.
It can be shown that for $0<p/m<1-2m/(1+a^2)$ the spacetime does not have an apparent horizon ---it would be at $\zeta_{ah}>\zeta_s$ and therefore outside the physical range for $\zeta$.

Defining a criticality parameter $\alpha$ according to
\be
\alpha \equiv \left[m\left(1-\frac{2m}{1+a^2}\right)-p\right]\,,
\ee
we can power expand the expression for the apparent horizon around the critical value $\alpha=0$:
\be
\zeta_{ah} = \zeta_s +\frac{a^2\left(1+a^2\right)}{m^2\left(a^2+a^4+2m\right)}  \alpha + O(\alpha^2)\,.
\ee
Finally, plugging this into Eq.~\eqref{eq:MassRatio} we obtain
\be
\frac{M(v,r_{ah})}{M(v,r\to\infty)} \propto (-\alpha)^\frac{a^2}{1+a^2}\,.
\ee
Therefore, we read off the critical exponent, as a function of the dilaton coupling,
\be
\gamma = \frac{a^2}{1+a^2}\,.
\ee
Of course, this could have been directly obtained from the result of section~\ref{sec:genericCSS}.
Thus, we find that the critical exponent can take any value between $0$ and $1$. For the heterotic coupling $a=1$ we recover our earlier result $\gamma=1/2$.

\subsection{Energy conditions on the total stress-energy tensor}
\label{sec:EnergyConds}

The various energy conditions (null, weak, strong and dominant) of the total stress-energy tensor can be expressed in terms of its eigenvectors and eigenvalues~\cite{Hawking:1973uf, Kuchar:1990vy}. It  turns out that the total stress-energy tensor supporting the solutions considered in this section possesses one timelike eigenvector $E^\mu_{0}$ and three spacelike eigenvectors $E^\mu_{1}$, $E^\mu_{2}$ and $E^\mu_{3}$. Adopting the nomenclature of Ref.~\cite{Hawking:1973uf}, this is a stress-energy tensor of type I. Explicitly, the eigenvectors are
\begin{subequations}
\bea
E^\mu_{(0)} &=& \frac{2 |a m \zeta|}{|a m \zeta| f(\zeta) - \sqrt{ \Theta(\zeta) }} \delta^\mu_v +\delta^\mu_r \,, \\
E^\mu_{(1)} &=& \frac{2 |a m \zeta|}{|a m \zeta| f(\zeta) + \sqrt{ \Theta(\zeta) }} \delta^\mu_v +\delta^\mu_r  \,, \\
E^\mu_{(2)} &=& \delta^\mu_\theta \,, \\
E^\mu_{(3)} &=& \delta^\mu_\phi \,, 
\eea
\end{subequations}
where we have defined 
\be
\Theta(\zeta) = a^2 m^2 \left(2- \zeta f(\zeta)\right)^2 + 2 \left(1+a^2\right)^2(1- m \zeta )p \,.
\ee
We note that $\Theta(\zeta)$ is always non-negative on account of $p>0$. 

The eigenvalues associated to these vectors are 
\begin{subequations}
\begin{align} \label{eq_lambda_0}
	\lambda_0 &=
	        \begin{aligned}[t]
	   & \frac{\left[ \left(1- \frac{m \zeta}{1+a^2}\right)^2  - \left( \frac{a m \zeta}{1+a^2} \right)^2 \right]f(\zeta) + \left(1 - \frac{m \zeta}{1+a^2}\right) (1- m \zeta) p \zeta }{8 \pi r^2 (1-m \zeta)^2} \\
	   & - \frac{  \frac{2a^2 m}{1+a^2} \left(1- \frac{2m \zeta}{1+a^2}\right) + (1-m \zeta)^\frac{2}{1+a^2} + \frac{|a m \zeta| \sqrt{\Theta(\zeta)}}{\left(1+a^2\right)^2} }{8 \pi r^2 (1-m \zeta)^2} \,,
	           \end{aligned} \\ \label{eq_lambda_1}
	 \lambda_1 &=    \lambda_0 + \frac{|a m \zeta| \sqrt{\Theta(\zeta)}}{4 \pi r^2 \left(1+a^2\right)^2 (1-m \zeta)^2}       \,, \\ \label{eq_lambda_2}
	\lambda_2 &= \lambda_3 = \frac{a^2 m \zeta \left\{ m \left[2-  \zeta f(\zeta) \right] + \left(1+a^2 \right) \left(1-m \zeta \right) p \zeta\right\} }{8 \pi r^2 \left(1+a^2\right)^2 \left(1- m \zeta\right)^2}  \,.
\end{align}
\end{subequations}
For this type of stress-energy tensor $- \lambda_0$ corresponds to the proper energy density while the $\lambda_i$ with $i=1,2,3$ correspond to the principal stresses.

The energy conditions are as follows. The null energy condition (NEC) requires $-\lambda_0  +\lambda_i\geq 0$ for $i=1,2,3$. The weak energy condition (WEC) moreover demands $-\lambda_0 \geq 0$. The strong energy condition (SEC) imposes the NEC and that $\sum_{i=1}^{3} \lambda_i \geq \lambda_0$. Finally, the dominant energy condition (DEC) requires the WEC and additionally that $-\lambda_0 -\lambda_i \geq 0$ for $i=1,2,3$.

To analyze the energy conditions, in many cases it suffices to estimate a maximum for the number of roots of the generalized polynomial that is obtained by doing a change of variable $\zeta \mapsto (1-x)/m$ and resorting to the generalized Descartes' rule (see Ref.~\cite{Jameson:2006}). Comparing this number with the behavior of the expressions defining the energy conditions one can then conclude about the existence or not of roots in the interval $x \in \left(0,1\right]$ (and therefore whether the inequalities are violated or satisfied). This strategy allowed us to fully map the parameter space according to the energy conditions, but only for $a\geq1$. For $a<1$ a similar analysis is not effective and we have therefore resorted to numerical evaluations of the energy conditions. We relegate the detailed computations to Appendix~\ref{App1} but the upshot of the analysis is summarized in Table~\ref{tab:EnergyConditions} below.

\renewcommand{\arraystretch}{1.5}
\begin{table}[htp]
\begin{center}
\begin{tabular}{c|c|c|c|c}
$a$         &         $m$          &        $p$        &   character of singularity   &   energy conditions \\
\hline
\hline
\multirow{3}{*}{$a>1$}    &   $2m\geq 1+a^2$   &               $p>0$                  &   spacelike   &   all satisfied            \\
\cline{2-5}
&   \multirow{2}{*}{$0<2m<1+a^2$}     &   $\frac{p}{m}\geq 1-\frac{2m}{1+a^2}$   &  spacelike or lightlike  &   all satisfied            \\
\cline{3-5}
&        &   $0<\frac{p}{m}<1-\frac{2m}{1+a^2}$    &  timelike  &   all violated              \\
\hline
$a=1$                             &   $m>0$   &   $p>0$   &   space-, light- or timelike   &   all satisfied   \\
\hline
\multirow{4}{*}{$a<1$}    &   $2m\geq 1+a^2$   &                $p>0$                 &   spacelike   &   see Figs.~\ref{fig:NEC}--\ref{fig:DEC}, *           \\
\cline{2-5}
&   \multirow{3}{*}{$0<2m<1+a^2$}     &   $\frac{p}{m}> 1-\frac{2m}{1+a^2}$   &  spacelike  &   see Figs.~\ref{fig:NEC}--\ref{fig:DEC}, *           \\
\cline{3-5}
&      &   $\frac{p}{m} = 1-\frac{2m}{1+a^2}$   &  lightlike  &   see Figs.~\ref{fig:NEC}--\ref{fig:DEC}           \\
\cline{3-5}
&        &   $0<\frac{p}{m}<1-\frac{2m}{1+a^2}$    &  timelike  &   see Figs.~\ref{fig:NEC}--\ref{fig:DEC}            \\
\end{tabular}
\end{center}
\vspace{-0.5cm}
\caption{Classification of the parameter space $(a,m,p)$ according to the character of singular surface and to the energy conditions. The cases marked with * indicate that the DEC is necessarily violated.\label{tab:EnergyConditions}}
\end{table}
%
%

\section{A time-dependent dyonic vacuum solution of the EMD system}
\label{sec:Dyonic}

So far we have focused our attention on continuously self-similar spacetimes and ---with the exception of section~\ref{sec:SSconditions}--- in particular those that are sourced by null fluids. In this section we instead abandon the assumption of CSS and look for source-free solutions.

We present below a novel vacuum (but non-static) dyonic solution of EMD theory, which is derived through a procedure we previously employed in Ref.~\cite{Aniceto:2017gtx}. The technique essentially amounts to taking a known static solution, promoting the mass and dilaton charge to be functions of advanced or retarded time while keeping the electromagnetic charges constant, and finally imposing constraints so that the field equations are satisfied without additional sources. This method is effective only for $a=1$, corresponding to the heterotic string value of the dilaton coupling, so this is the case we will restrict to.

Source-free dynamical solutions of the EMD system (with $a=1$ and a purely electric Maxwell field) were obtained in Refs.~\cite{Gueven:1996zm, Aniceto:2017gtx} by taking different routes. In any case, the end result were the following expressions for the line element and matter fields,
\bea
 ds^2 &=& -\left(1-\frac{2M(v)}{r}-2\epsilon D'(v)\right) dv^2 - 2\, \epsilon\, dv dr + r^2 \left(1-\frac{2D(v)}{r}\right) d\Omega^2\,, \nn\\
F &=& -\frac{Q}{r^2} dv\wedge dr\,,\\
e^{2\Phi} &=& 1-\frac{2D(v)}{r}\,. \nn
\eea
These expressions were supplemented with the `vacuum' constraints,
\be
2M(v)D(v) = Q^2= {\rm const.}\,, \qquad\qquad
2 D(v)^3 D''(v)+\epsilon\, Q^2 D'(v)=0\,.
\label{eq:vaccond}
\ee
Here $\epsilon$ is just a sign, accounting for the situation in which $v$ is a retarded ($\epsilon=+1$) or an advanced ($\epsilon=-1$) Eddington-Finkelstein time coordinate.
Note that if one sets the (electric) charge to zero, the vacuum constraints imply that $D(v)$ is linear and $M(v)$ vanishes. Thus, Roberts' self-similar solution~\eqref{eq:Roberts} is recovered.

Time-dependent {\em dyonic} solutions were generated in~\cite{Aniceto:2017gtx} by taking the above purely electric dynamical solution, regarded as a solution of Einstein-Maxwell-axion-dilaton system with a vanishing axion, and applying S-duality. The outcome of this exercise is a spacetime with the same line element, but the electromagnetic field strength and the dilaton are transformed, while the axion is nonvanishing.
In this case there is both an electric and a magnetic charge, $Q_e$ and $Q_m$, and the total charge appearing in the metric components is related to $Q_e$ and $Q_m$ through
\be
Q^2=Q_e^2+Q_m^2\,.
\label{eq:QQeQm_plus}
\ee
The same vacuum conditions~\eqref{eq:vaccond} apply, but now it is the total electromagnetic charge that appears in the constraints.


An analogous class of dynamical dyonic solutions but without the axion ---and therefore in EMD theory--- can be obtained as a generalization of the static dyonic solutions of Kallosh et al.~\cite{Kallosh:1992ii}. This is a four-parameter family of solutions to EMD theory defined by a mass $M$, an electric charge $Q_e$, a magnetic charge $Q_m$, and a dilaton charge $\Sigma$. Applying the procedure described in~\cite{Aniceto:2017gtx} to these static dyonic solutions one obtains
\bea
ds^2 &=&-\left(1-\frac{ 2 M(v)}{r + \Sigma(v)} + \frac{ 2Q_e^2}{r^2 - \Sigma(v)^2}\right) dv^2 - 2\, \epsilon\, dv dr + \left(r^2-\Sigma(v)^2\right) d\Omega^2\,,\nn\\
F &=& -\frac{Q_e}{(r-\Sigma(v))^2} dv\wedge dr + Q_m \sin\theta\, d\theta\wedge d\varphi\,,\\
 e^{-2\Phi} &=& \frac{r-\Sigma (v)}{r+\Sigma (v)}\,, \nn
\eea
with the constraints
\be
2M(v)\Sigma(v) = Q_m^2-Q_e^2= {\rm const.}\,, \qquad\qquad
2 \Sigma(v)^3 \Sigma''(v)+\epsilon\, (Q_m^2-Q_e^2) \Sigma'(v)=0\,.
\ee

We can make contact with our previous solutions by shifting $r\to r+\Sigma(v)$ and setting $\Sigma(v) = -D(v)$, which yields
\bea
ds^2 &=&-\left(1-\frac{ 2 M(v)}{r} - 2\epsilon D'(v) + \frac{2Q_m^2}{r(r-2D(v))} \right) dv^2 - 2\, \epsilon\, dv dr + r^2\left(1-\frac{2D(v)}{r}\right) d\Omega^2\,,\nn\\
F &=& -\frac{Q_e}{r^2} dv\wedge dr + Q_m \sin\theta\, d\theta\wedge d\varphi\,, \label{eq:dyonicE}\\
 e^{2\phi} &=& \left(1-\frac{2D(v)}{r}\right)\,. \nn
\eea
Interestingly, the vacuum constraints become the same as~\eqref{eq:vaccond}, but now with
\be
Q^2 = Q_e^2 - Q_m^2\,.
\label{eq:QQeQm_minus}
\ee
Note the difference in sign with respect to~\eqref{eq:QQeQm_plus}.

Using the algebraic relation above (and again shifting the radial coordinate) we can write this solution in 
the alternative form
\bea
ds^2 &=&-\left(1-\frac{2 M(v)}{r} - 2\epsilon D'(v) + \frac{2Q_e^2}{r(r+2D(v))} \right) dv^2 - 2\, \epsilon\, dv dr + r^2\left(1+\frac{2D(v)}{r}\right) d\Omega^2\,,\nn\\
F &=& -\frac{ Q_e}{(r+2D(v))^2} dv\wedge dr + Q_m \sin\theta\, d\theta\wedge d\varphi\,, \label{eq:dyonicM}\\
 e^{-2\phi} &=& \left(1+\frac{2D(v)}{r}\right)\,. \nn
\eea
The two alternative forms~\eqref{eq:dyonicE} and~\eqref{eq:dyonicM} show that both the purely electric and purely magnetic solutions naturally coincide with the dilatonic black holes of~\cite{Gibbons:1987ps,Garfinkle:1990qj} in the static limit.


These dyonic solutions can never be self-similar because the constant (nonvanishing) electromagnetic charge introduces a length scale, thus destroying scale-invariance. Nevertheless, those solutions with $Q_e = \pm Q_m$ come close as in this case the vacuum constraints impose
\be
D(v)=\sigma v \qquad \text{with } \sigma = const.\,, \qquad\qquad
M(v)=0\,.
\ee
This is exactly what we required to retrieve Roberts' self-similar solution in the neutral case, but now there is a fixed scale ---namely $|Q_e| = |Q_m|$--- that breaks self-similarity\footnote{In practice, this is reflected by the fact that the metric component $g_{vv}$ is not a function of the ration $v/r$ only.},
\bea
ds^2 &=&-\left(1- 2\epsilon \sigma + \frac{2Q_m^2}{r^2(1-2\sigma v/r)} \right) dv^2 - 2\, \epsilon\, dv dr + r^2\left(1-\frac{2\sigma v}{r}\right) d\Omega^2\,,\nn\\
F &=& -\frac{Q_m}{r^2} dv\wedge dr + Q_m \sin\theta\, d\theta\wedge d\varphi\,,\\
 e^{2\phi} &=& \left(1-\frac{2\sigma v}{r}\right)\,. \nn
\eea
Obviously, for vanishing electromagnetic charges one recovers Roberts' spacetime~\eqref{eq:Roberts}.

\section{Conclusion}
\label{sec:Conc}

In conclusion, we have obtained a two-parameter family of continuous self-similar solutions to Einstein-Maxwell-dilaton theory supported by charged null fluids, in spherical symmetry. These spacetimes were then used to study critical behavior analytically, within this understudied (but relevant) matter model. The critical exponent we computed generally differs from the usual $1/2$ value, and in fact can take any value between $0$ and $1$, depending on the dilaton coupling $a$ that defines the theory, but for the specific heterotic string value ($a=1$) one recovers the typical result of $1/2$.

On a somewhat peripheral note, we also presented a new class of time-dependent dyonic exact solutions to this theory when $a=1$. These source-free spacetimes are not self-similar, and they represent the counterpart of the dyonic solutions obtained in Ref.~\cite{Aniceto:2017gtx} for the Einstein-Maxwell-axion-dilaton theory.

We would like to end with some remarks. The first one concerns weak cosmic censorship. Upon matching our CCS solutions sourced by null fluids with flat space, we obtained global solutions that describe the formation of a naked singularity (in the subcritical case). Moreover, this can be done without ever violating any energy conditions when $a\leq1$. Nevertheless, the restriction imposed by spherical symmetry and self-similarity make these spacetimes highly nongeneric.

Secondly, we should leave a word of caution. The critical behavior we have tackled with analytic methods in this paper specifically refers to the observation that we obtained power-law scaling for the (normalized) mass of the black holes formed, with a universal exponent. In our case there is really no emergent symmetry near criticality because we imposed continuous self-similarity from the outset to obtain the entire family of solutions, not just the critical solution. Furthermore, we did not impose analyticity on our global solutions, nor have we showed that their spectrum of perturbations possesses a single growing mode, as happens to be the case for the massless scalar field~\cite{Gundlach:1996eg}, so our solutions are not critical in this strict sense.

Finally, we found that energy conditions are violated in certain regions of the parameter space. This typically (but not always) occurs at some surface $\zeta=\zeta_v$. Unfortunately, it is generally difficult to obtain an exact expression for $\zeta_v$, but if it is the case that such a surface is spacelike, one might use the techniques of \cite{Ori:1991,Chatterjee:2015cyv,Creelman:2016laj} to glue the original solution (with infalling null matter) onto an outgoing version of itself, in order to obtain a global spacetime free from energy condition violations, and which describes a bouncing null fluid.

\acknowledgments

We thank Roberto Emparan and Carsten Gundlach for useful discussions. 
PA acknowledges financial support from Funda\c{c}\~ao para a Ci\^encia e Tecnologia (FCT) through the LisMath fellowship PD/BD/128415/2017.
JVR acknowledges financial support from the European Union's Horizon 2020 research and innovation programme under ERC Advanced Grant GravBHs-692951.
Funding for this work was partially provided by the Spanish MINECO under projects FPA-2016-76005-C2-2-P and MDM-2014-0369 of ICCUB (Unidad de Excelencia `Mar\'ia de Maeztu').

\appendix
\section{Analysis of the energy conditions}
\label{App1}

In this appendix we collect the computations required to assess the classical energy conditions for the CSS solutions supported by null fluids presented in section~\ref{sec:CSSsourcedSols}.

\paragraph{Null energy condition}

The condition $-\lambda_0 + \lambda_1 \geq 0$ is always satisfied. This can be seen directly from Eq.~\eqref{eq_lambda_1} since $\Theta(\zeta)\geq0$. On the other hand, Eqs.~\eqref{eq_lambda_0} and~\eqref{eq_lambda_2} tell us that 
\be \label{eq_NEC}
 \lambda_2 - \lambda_0 = \frac{2a^2 m \left( a^2+x\right) + \left(1+a^2\right)^2 \left[ x^\frac{2}{1+a^2} - \frac{p}{m} x^2 \left(1-x\right) \right]  -  \left(a^2 +x\right)^2 f\left(\frac{1-x}{m}\right)  + |a|  \left(1-x\right)\sqrt{\Theta\left(\frac{1-x}{m}\right)  } }{8 \pi r^2 \left(1+a^2\right)^2  x^2 } \,,
\ee
where we have made the change of coordinate $\zeta \mapsto \left(1-x\right)/m$. We now separate our analysis in two different cases. First, when we have a non-timelike singularity, i.e., $p/m\geq 1-2m/(1+a^2)$ and $2m \leq 1+a^2$ or $p>0$ and $2m>1+a^2$ we obtain that the NEC is necessarily satisfied if 
\be \label{eq_NEC_polynomial}
a^4\left[\frac{p}{m} - \left(1-\frac{2m}{1+a^2}\right)\right] + \left(1+a^2\right)^2 x^\frac{2}{1+a^2}  - a^2 c_1 x - c_2 x^2 + a^2 \left(a^2+2\right) \frac{p}{m} x^3 \geq 0 \,,
\ee
with $c_1 = 2+ 2m\left(a^2-1\right)/\left(1+a^2\right) + \left(a^2-2\right)p/m$ and $c_2 = 1+ 2a^2m/(1+a^2) + \left(4+a^2\right)a^2 p/m$. Resorting to the results of Ref.~\cite{Jameson:2006}, we note that for all $a \geq 1$ the (ordered) coefficients of this generalized polynomial change sign twice, which implies there are at most two zeroes for $x >0$. The generalized polynomial vanishes at $x=1$ and is always positive at $x=0$ for a spacelike singularity. Additionally, we have that its first derivative with respect to $x$ is negative at $x=1$. Hence, a root of the polynomial for $x \in \left(0,1\right]$ necessarily occurs only at $x=1$, from which we conclude that in this case the NEC is satisfied. When the singularity is lightlike, i.e., $p/m = 1- 2m/(1+a^2)$ we have that $x=0$ is also a zero of Eq.~\eqref{eq_NEC}. In this case we must then compute the value of the first derivative with respect to $x$ at that point, finding it is always positive for $a\geq 1$. Hence, for a lightlike singularity, the NEC is also satisfied when $a\geq 1$.

For a timelike singularity, i.e., $p/m< 1-2m/(1+a^2)$ with $2m<1+a^2$, we have that for the NEC to be satisfied it is necessary that
\be
a^2\left(a^2-1\right) \left[\frac{p}{m} - \left(1-\frac{2m}{1+a^2}\right) \right] \geq 0 \,.
\ee
This follows immediately from the numerator of~\eqref{eq_NEC} evaluated at $x=0$, and we see it is violated when $a>1$. On the other hand when $a=1$ we find that it is always satisfied.

\begin{figure}[t]
\centering
\includegraphics[width=8cm]{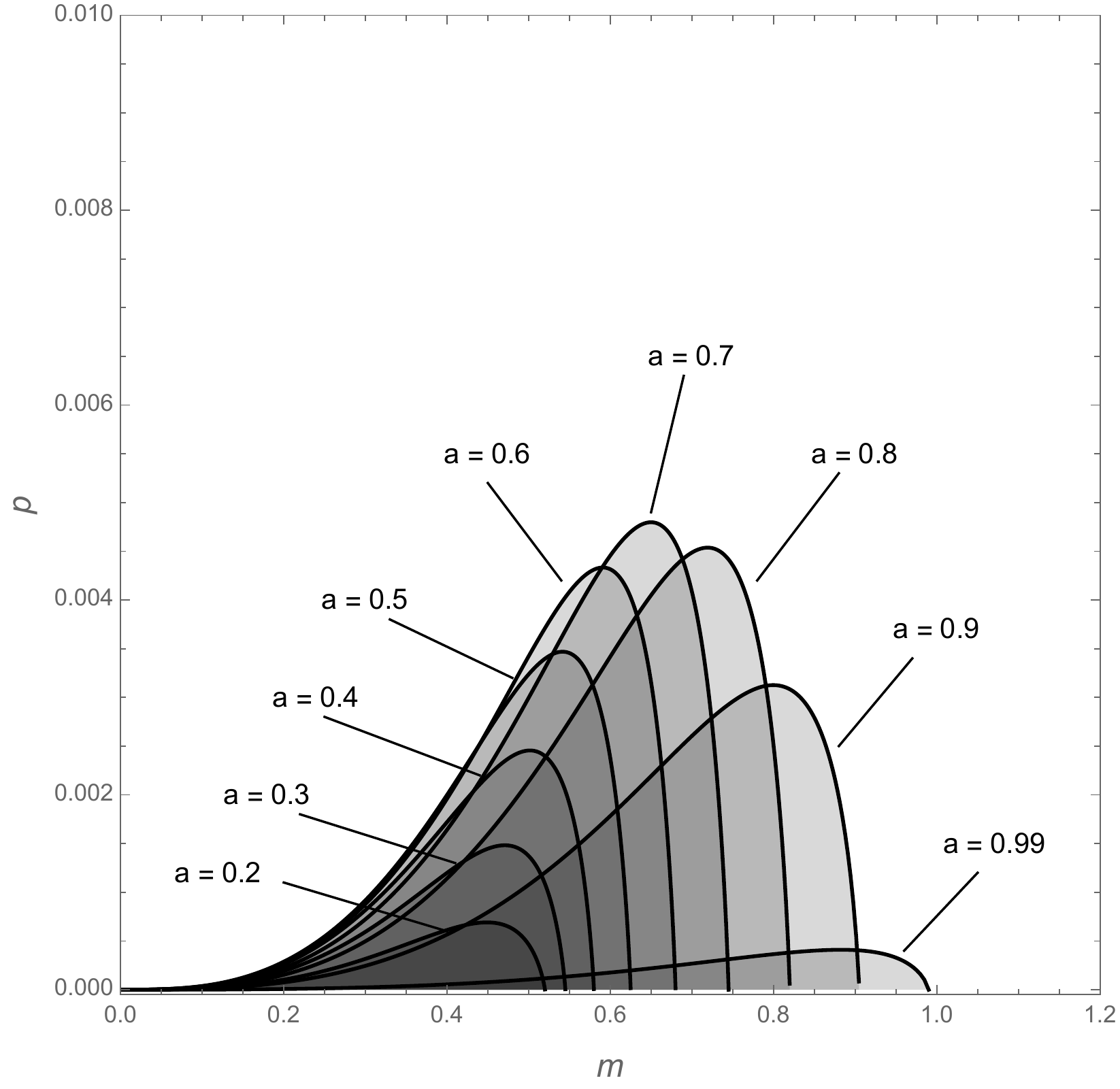}
\caption{The shaded areas indicate regions in the parameter space where the NEC (as well as the SEC) is violated, for various values of the dilaton coupling $0<a<1$. Note the smallness of vertical scale.}
\label{fig:NEC}
\end{figure}

The analysis so far was sufficient to fully characterize the parameter space according to conformity or violation of the null energy condition when $a\geq1$. However, the approach used is inconclusive when $a<1$, because in this case the relevant generalized polynomial may have an additional root in the interval $\left(0,1\right]$. 
Hence, we have resorted to numerical methods to assess the validity of the NEC, employing a similar strategy as that used in Ref.~\cite{Rocha:2017uwx}. In essence, this amounts to numerically searching for double roots of the generalized polynomial, which determine the boundary between compliance and violation of the energy condition.
The results are presented in Fig.~\ref{fig:NEC} and indicate that the NEC is satisfied almost over the entire parameter space when $0<a<1$, being violated only for very small values of the parameter $p$, namely $p\lsim0.005$.

\paragraph{Weak energy condition} Since the WEC implies the NEC we only need to evaluate which of the previous results also satisfy the condition $-\lambda_0 \geq 0$. This condition is satisfied if
\be \label{eq_WEC_polynomial}
a^2 \left(a^2-1\right) \left[ \frac{p}{m} - \left(1- \frac{2m}{1+a^2}\right)\right] + \left(1+a^2\right)^2 x^\frac{2}{1+a^2} - 2a^2 c_1x + b_2 x^2 + \frac{2a^2p}{m} x^3  \geq 0 \,,
\ee 
with $c_1$ defined as in Eq.~\eqref{eq_NEC_polynomial} and $b_2 =\left(a^2-1\right)\left[1+ a^2 2m/(1+a^2)+ a^2 p/m  \right]  - 4 a^2 p/m$. This generalized polynomial features two sign changes and has a root at $x=1$. Moreover, its derivative w.r.t. $x$ at $x=1$ is always negative and the polynomial at $x=0$ is always positive for a spacelike singularity. Hence, for spacelike singularities the WEC is always satisfied for $a \geq 1$. For lightlike singularities $x=0$ is also a root of the polynomial and the derivative with respect to $x$ at that point is always positive for $a\geq 1$, from which we conclude that the WEC is also satisfied for lightlike singularities when $a \geq 1$.

When we have a timelike singularity, we already know from the NEC that the WEC is violated for $a>1$. When $a=1$ we once again obtain that $-\lambda_0$ is always non-negative. Thus, for a timelike singularity the WEC is satisfied when $a=1$.

\begin{figure}[t]
\centering
\includegraphics[width=8cm]{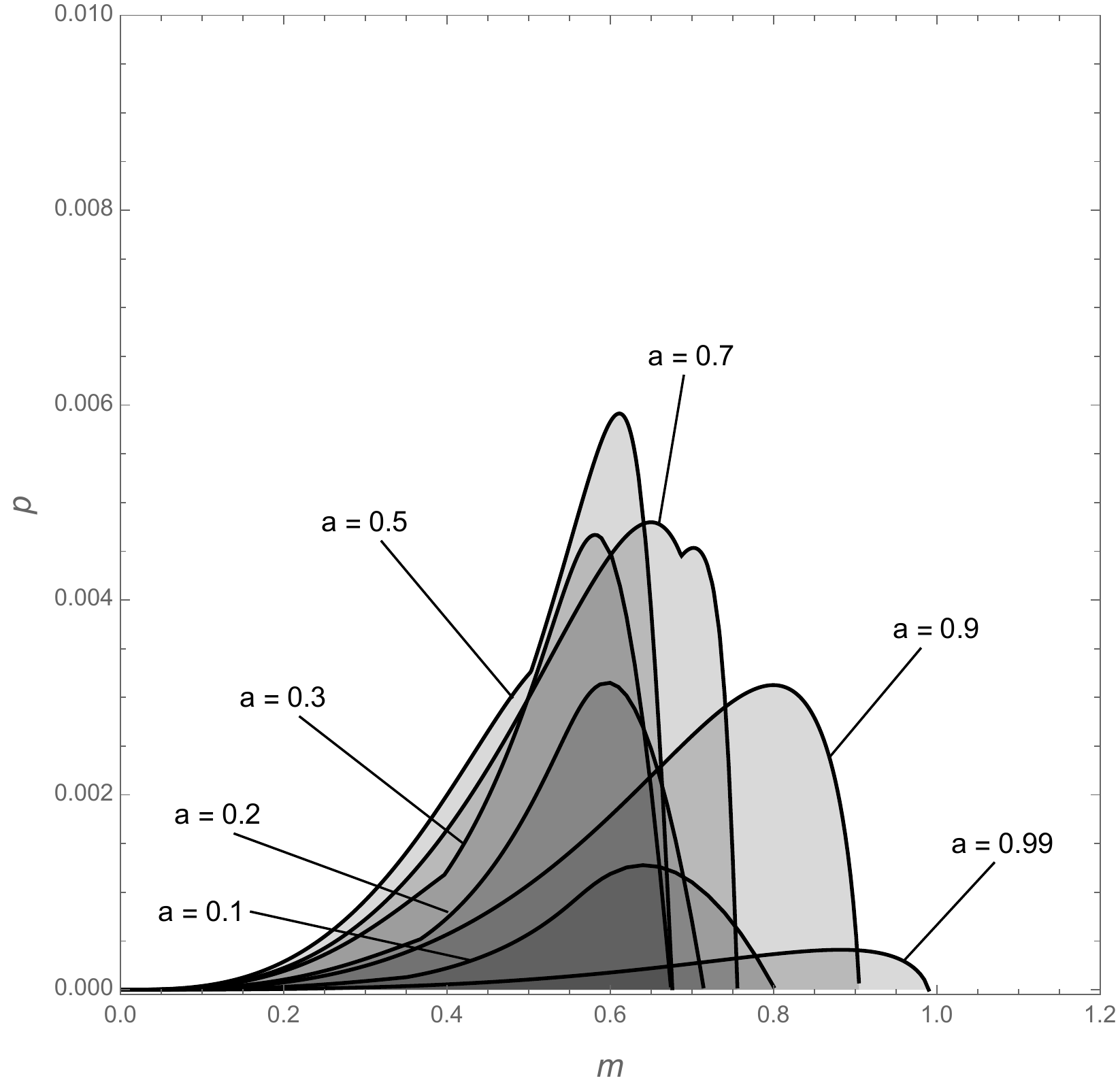}
\caption{The shaded areas indicate regions in the parameter space where the WEC is violated, for various values of the dilaton coupling $0<a<1$. For $a\gsim0.8$ these curves coincide with those of Fig.~\ref{fig:NEC} because the additional condition $-\lambda_0\geq0$ imposes no further constraints.  For $a\lsim0.8$ the WEC is more restrictive than the NEC and the violation regions consequently become more intricate. Note the smallness of vertical scale.}
\label{fig:WEC}
\end{figure}

For the same reasons as above, this analysis is inconclusive when $a<1$, so we resort again to numerical evaluations of the WEC.
The results, displayed in Fig.~\ref{fig:WEC}, show that the WEC is also satisfied over most of the parameter space when $0<a<1$, being violated only for very small values of the parameter $p$, namely $p\lsim0.006$.

\paragraph{Dominant energy condition}  The inequality $-\lambda_0 - \lambda_1 \geq 0$ is equivalent to Eq.~\eqref{eq_WEC_polynomial}, and therefore it is satisfied under the same conditions. For the condition $-\lambda_0 - \lambda_2\geq0$ we start by noting that 
\be
\Theta\left(\frac{1-x}{m}\right) \geq \Gamma(x)^2 \equiv \frac{a^2 \left[ 2m^2 \left(1+a^2 x\right) - \left(1+a^2\right) \left(1-x\right) \left(m - p + p x\right)    \right]^2 }{\left(1+a^2\right)^2 m^2} \,.
\ee
Using this, we can minorate $\sqrt{\Theta} \geq \sqrt{\Gamma(x)^2} \geq \Gamma(x)$ and conclude that a sufficient condition for the inequality $-\lambda_0 - \lambda_2 \geq 0$ to be satisfied is
\be \label{eq_DEC_polynomial}
\left(1+a^2\right)^2 x^\frac{2}{1+a^2} - \frac{a^2d_1}{1+a^2}  x + \frac{a^2-1}{1+a^2} d_2 x^2 - \frac{d_3 d_4  }{1+a^2} x^3 + d_4 \left(1-x\right)^3  \left[\frac{p}{m}-\left(1-\frac{2m}{1+a^2}\right)\right] \geq 0 \,,
\ee
with $d_1= 1+3 a^4 +4a^2 +2m\left(1-a^2\right)$, $d_2 = 1+3a^4 +4a^2 \left(1-m\right)$, $d_3= \left(1+a^2-2m\right)$ and $d_4 = a^2 \left(a^2-1\right)$. When we have a spacelike or lightlike singularity the last term is always non-negative so we ignore it to simplify our analysis. Consequently, we end up with a generalized polynomial whose (ordered) coefficients change sign at least twice and at most three times. The polynomial has a root at $x=0$ and another one of multiplicity two at $x=1$ and with a positive second derivative at this point for $a>1$. Furthermore, for $a>1$ its first derivative is always positive at $x=0$. Moreover, when $a=1$, the left hand side of Eq.~\eqref{eq_DEC_polynomial} vanishes. We then conclude that, for $a\geq 1$, the DEC is satisfied for spacelike and lightlike singularities. 

For timelike singularities, since we know that the WEC is violated for $a>1$, it suffices to check what happens when $a=1$. Computing the values of $-\lambda_0 -\lambda_1$ and $-\lambda_0 -\lambda_2$ when $a=1$ we find that both conditions are satisfied in this case. Hence, for timelike singularities, the DEC is satisfied when $a=1$. 

\begin{figure}[t]
\centering
\includegraphics[width=8cm]{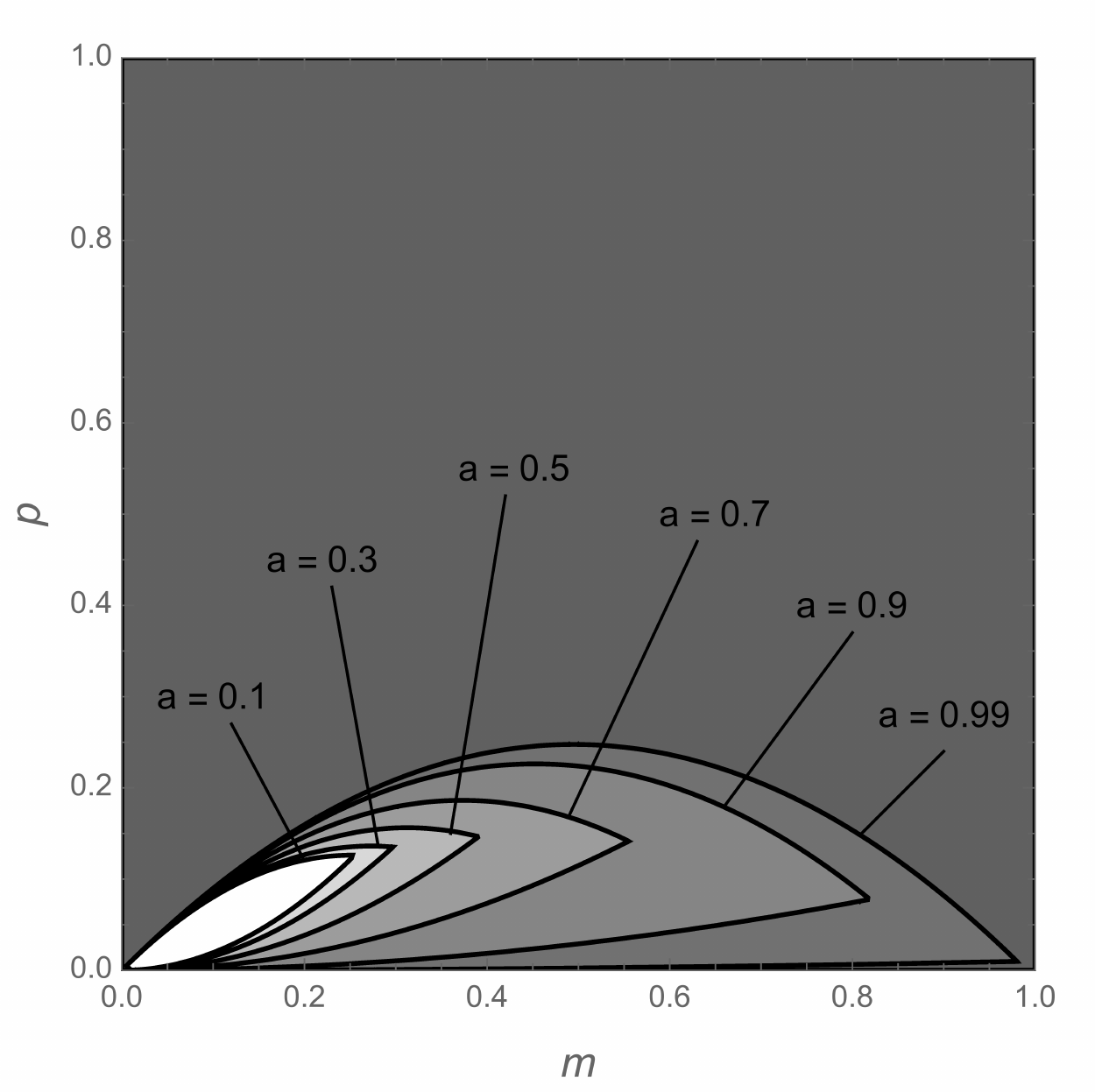}
\caption{The shaded areas (exterior to the almond shaped areas) indicate regions in the parameter space where the DEC is violated, for various values of the dilaton coupling $0<a<1$. These regions are not optimal, in the sense that they were obtained as sufficient but not necessary conditions for the DEC to be violated.}
\label{fig:DEC}
\end{figure}

Once again, this analysis is inconclusive when $a<1$. Contrarily to the NEC and WEC studies, we have not been able to use numerical methods to determine the regions in parameter space where the DEC is violated, due to very steep gradients featured by the generalized polynomial that defines the DEC. The best we have managed to obtain are sufficient conditions for the DEC to be violated:
\be
\frac{p}{m} > 1 - \frac{2m}{1+a^2} \,, \qquad
\frac{p}{m} < 2m\frac{1-a^2}{(1+a^2)^2} \,.
\ee
The first constraint (which is equivalent to the singularity being spacelike) arises from the quantity $-\lambda_0 - \lambda_1$ evaluating to negative values at $\zeta=\zeta_s=1/m$, while the second condition stems from the vanishing of $-\lambda_0 - \lambda_1$ at $\zeta=0$, but with negative derivative.
These constraints are plotted in Fig.~\ref{fig:DEC} and indicate that the DEC is violated almost over the entire parameter space when $0<a<1$, leaving only a small almond-shaped region where it may be satisfied.

\paragraph{Strong energy condition} For the SEC we find that the condition $\lambda_1 + \lambda_2 + \lambda_3 - \lambda_0\geq0$ is always satisfied for all $x\in \left[0,1\right]$. Hence, the requirements for the SEC to be satisfied are equivalent to the requirements for the NEC to be satisfied.



\end{document}